\documentclass[%
preprint,
 amsmath,amssymb,
 aps,
pra,
]{revtex4-2}

\usepackage{subfigure}
\usepackage{graphicx}
\usepackage{dcolumn}
\usepackage{bm}
\usepackage{amsmath,amsfonts,amssymb}

\begin{document}

\title{Optimal shortcuts of Stimulated Raman Adiabatic Passage in the presence of dissipation}

\author{Dionisis Stefanatos}
\email{dionisis@post.harvard.edu}

\author{Emmanuel Paspalakis}

\affiliation{Materials Science Department, School of Natural Sciences, University of Patras, Patras 26504, Greece}


\begin{abstract}
We use optimal control theory to obtain shortcuts to adiabaticity which maximize population transfer in a three-level STIRAP system, for a given finite duration of the process and specified dissipation rate at the intermediate state. We fix the sum of the intensities of the pump and Stokes pulses and use the mixing angle as the sole control variable. We determine the optimal variation of this angle and reveal the role of singular arc in the optimal trajectory, in order to minimize the effect of dissipation.
\end{abstract}


\maketitle

\section{Introduction}

Stimulated Raman adiabatic passage (STIRAP) is a renowned method for robust population transfer between two non-directly connected states of a three-level quantum system \cite{Bergmann98,Kral07,Vitanov17,Sola18,Bergmann19,Vitanov97,Kobrak98}. The transfer between the initial state $|1\rangle$ and the target state $|3\rangle$ is accomplished via a lossy intermediate state $|2\rangle$, which is coupled to the other two through the pump and Stokes laser fields, respectively. The trademark of STIRAP is the counterintuitive order of application of the Stokes and pump fields, where the former precedes the latter. As a consequence, a coherent superposition of states $|1\rangle$ and $|3\rangle$ is built and, as time passes, it is transformed from the starting state to the target state. If the Stokes and pump control fields are slowly (adiabatically) varied, the lossy mid-state $|2\rangle$ is barely populated, and perfect population transfer is eventually realized. The main advantage of STIRAP is its robustness to modest imperfections in the experimental parameters, and for this reason has found application in many areas of contemporary physics \cite{Bergmann19}, for example optical waveguides \cite{Paspalakis06,Dreisow09}, nitrogen-vacancy centers in diamond \cite{Golter14}, matter waves \cite{Menchon16}, and superconducting quantum circuits \cite{Kumar16}.


In an effort to increase STIRAP efficiency, optimal control \cite{Bryson75} has been exploited, as in many other quantum applications \cite{StefanatosReview,Boscain21}. Although it was initially realised that the counterintuitive pulses with zero initial values cannot make up the optimal solution when the objective is to maximize the target state population \cite{Band94}, it was soon after shown with numerical optimal control that the counterintuitive pulse sequence becomes optimal as long as a penalty term punishing the population of the lossy intermediate state is incorporated in the cost function \cite{Sola99,Kis02,Kumar11}. In another work \cite{Boscain02}, omitting dissipation and setting as goal the minimization of the transfer time to the target state or of the fluence of the applied fields, the optimal solution was found to be the intuitive pulse sequence, where the pump pulse precedes the Stokes pulse. In a subsequent study \cite{Rat12}, where dissipation was taken into account, by exploiting the methods developed in Refs. \cite{Khaneja03,Stefanatos04,Stefanatos05} (see also Ref. \cite{Khaneja17}) with bounded Stokes and unbounded pump pulses, an optimal solution resembling the counterintuitive pulse sequence was obtained, with a constant Stokes pulse and a pump pulse which is much smaller initially and becomes much larger towards the end. In another investigation \cite{Assemat12}, where dissipation is also taken into consideration, the optimal solution was found to be a counterintuitive pulse sequence with nonzero Stokes and pump pulses at the initial and final times, respectively, using again a cost function penalizing the occupation of the mid-state $|2\rangle$. Finally, Ref. \cite{Dalessandro20} studied the minimum transfer time problem in the absence of dissipation. In parallel with the previously discussed studies, which exploit analytical and numerical optimal control in order to improve population transfer in the STIRAP system, different variations of shortcuts to adiabaticity \cite{STA19} have also been used to tackle the same problem \cite{Demirplak05,Chen10b,Giannelli14,Masuda15,Li16,Clerk16,Kolbl19,Dridi20,Petiziol20,Blekos20,Zheng22,Dogra22}.

In this study we use optimal control theory to find the control fields which maximize the population transfer from state $|1\rangle$ to state $|3\rangle$ for a finite prescribed duration $T$, without imposing any penalty on the population of the mid-state $|2\rangle$. We formulate the optimal control problem by fixing the sum of intensities of the pump and Stokes pulses, so the total field amplitude is constant and the only parameter to be controlled is the mixing angle of the two fields. We also transform the system to the basis composed of the dark and bright states, as well as the intermediate state $|2\rangle$. In the transformed system, the time derivative of the mixing angle arises naturally as the control function. If we do not impose any restriction on this control variable, we find the optimal solution to has the general bang-singular-bang form, where the initial and final bangs correspond to instantaneous rotations of the mixing angle. If we require the derivative of the mixing angle to be nonnegative, in order to conform with shortcuts to adiabaticity so the total field evolves from the Stokes field initially to the pump field finally, then we find the optimal pulse sequence to has the bang-off-bang form for shorter durations and the bang-off-singular-off-bang form for longer ones. Note that the off pulses correspond to intervals where the mixing angle is kept constant. We show that the optimal solution is symmetric around the central time, so the initial and final bang pulses have the same strength while the off pulses have the same duration. Regarding the midway singular part of the trajectory, we derive singular control in terms of a feedback law and also obtain an expression for the singular surface. We explain that it is this singular control which exploits longer available durations in order to find optimal paths along which the effect of dissipation is minimized, achieving thus larger population transfer. We also discuss how the optimal solution can be further integrated with shortcuts to adiabaticity, as well as how the formulation of the problem in the transformed basis can take advantage of the spins to springs mapping. Note that in our recent work \cite{Stefanatos21a} we solved the same problem for the special case where dissipation is much larger than the amplitude of the applied fields, so the intermediate state $|2\rangle$ can be adiabatically eliminated from the equations, leading thus to a simpler two-dimensional problem. Here on the contrary we solve the full three-dimensional optimal control problem.

In the next section we formulate the optimal control problem, while in section \ref{sec:sol0} we derive as a warm up example the minimum-time optimal solution in the absence of dissipation. In section \ref{sec:solG} we solve the optimal control problem in the presence of dissipation and under the restriction of non-decreasing mixing angle. In section \ref{sec:example} we present a numerical example and discuss various aspects of the optimal solution, while the results of the present study are summarized in section \ref{sec:conclusion}.



\section{The optimal control problem}

\label{sec:system}

The probability amplitudes $C_1, C_2, C_3$ corresponding to states $|1\rangle, |2\rangle, |3\rangle$ obey, in the case of one-photon and two-photon resonance, the following Schr\"{o}dinger equation
\begin{equation}
\label{Schrodinger}
i
\left(\begin{array}{c}
    \dot{C}_1  \\
    \dot{C}_2 \\
    \dot{C}_3
\end{array}\right)
=
\frac{1}{2}
\left(\begin{array}{ccc}
    0 & \Omega_p(t) & 0 \\
    \Omega_p(t) & -i\Gamma & \Omega_s(t) \\
    0 & \Omega_s(t) & 0
  \end{array}\right)
\left(\begin{array}{c}
    C_1  \\
    C_2 \\
    C_3
\end{array}\right).
\end{equation}
with $\Omega_p(t), \Omega_s(t)$ being the Rabi frequencies corresponding to the pump and Stokes pulses, and $\Gamma$ the rate at which population dissipates from state $|2\rangle$. At the initial time $t=0$ only state $|1\rangle$ is populated
\begin{equation}
\label{initial_conditions}
C_1(0)=1, \quad C_2(0)=C_3(0)=0,
\end{equation}
and the objective is to obtain the controls pulses $\Omega_p(t), \Omega_s(t)$ maximizing the final population of state $|3\rangle$, $|C_3(T)|^2$, for a specified duration $T$.

With the aim to render this optimization problem well posed, we set a bound on the pulse amplitudes via the following condition
\begin{equation}
\label{bounded_fluence}
\Omega^2_p(t)+\Omega^2_s(t)=\Omega^2_0\quad\mbox{constant},
\end{equation}
which corresponds to having a total field with constant amplitude in the equivalent two-level picture. Under this constraint, the only control variable left is the direction of the total field, governed by the
mixing angle
\begin{equation}
\label{mixing_angle}
\tan{\theta(t)}=\frac{\Omega_p(t)}{\Omega_s(t)}.
\end{equation}
We can get rid of complex quantities if we use the following state variables \cite{Boscain02}
\begin{equation}
\label{transform1}
Z=C_1,\quad Y=-iC_2,\quad X=-C_3,
\end{equation}
so system (\ref{Schrodinger}) becomes
\begin{equation}
\label{XYZ}
\left(\begin{array}{c}
    \dot{Z}  \\
    \dot{Y} \\
    \dot{X}
\end{array}\right)
=
\frac{1}{2}
\left(\begin{array}{ccc}
    0 & \sin\theta(t) & 0 \\
    -\sin\theta(t) & -\Gamma & \cos\theta(t) \\
    0 & -\cos\theta(t) & 0
  \end{array}\right)
\left(\begin{array}{c}
    Z  \\
    Y \\
    X
\end{array}\right),
\end{equation}
where we have also normalized time as $\Omega_0t$ and dissipation as $\Gamma/\Omega_0$.
Note that $Z^2=|C_1|^2, Y^2=|C_2|^2, X^2=|C_3|^2$, thus initially the Bloch vector is aligned with the north pole and the goal is to maximize $X^2(T)=|C_3(T)|^2$. Additionally, observe from this equation that for $\theta=0$ the total filed vector is also aligned with the north pole and coincides with the Stokes field, while for $\theta=\pi/2$ is aligned with the $X$-axis and coincides with the pump field.

In order to solve the optimal control problem we will not use the above $XYZ$ reference frame but instead consider the transformation \cite{Ivanov05}
\begin{eqnarray}
z&=&Z\cos\theta+X\sin\theta,\label{t_z}\\
y&=&Y,\label{t_y}\\
x&=&Z\sin\theta-X\cos\theta,\label{t_x}
\end{eqnarray}
where $z, x$ correspond to the dark and bright states of the quantum system, respectively, while $y$ still represents the intermediate state $|2\rangle$. The transformed $xyz$ system is found to be
\begin{equation}
\label{xyz}
\left(\begin{array}{c}
    \dot{z}  \\
    \dot{y} \\
    \dot{x}
\end{array}\right)
=
\left(\begin{array}{ccc}
    0 & 0 & -u(t) \\\noalign{\vskip3pt}
    0 & -\frac{\Gamma}{2} & -\frac{1}{2} \\\noalign{\vskip3pt}
    u(t) & \frac{1}{2} & 0
  \end{array}\right)
\left(\begin{array}{c}
    z  \\
    y \\
    x
\end{array}\right),
\end{equation}
where the naturally arising control function $u(t)$ is the derivative of the mixing angle
\begin{equation}
\label{theta}
\dot{\theta}=u(t).
\end{equation}
On the mixing angle $\theta$ we apply the boundary conditions
\begin{equation}
\label{b_theta}
\theta(0)=0,\quad\theta(T)=\frac{\pi}{2},
\end{equation}
so in the original $XYZ$ frame the total field is aligned at the beginning with the north pole (where also lies the initial Bloch vector) and at the end with $X$-axis, mimicking thus  shortcuts to adiabaticity. Using the initial conditions for $X, Y, Z$ and the boundary conditions for $\theta$, we find from Eqs. (\ref{t_z})-(\ref{t_x}) the initial conditions for the transformed variables $x,y,z$
\begin{equation}
\label{starting point}
x(0)=y(0)=0, \quad z(0)=1,
\end{equation}
as well as the quantity to be maximized, $z(T)=X(T)$. The problem that we study in this paper is to find the optimal control $u(t)$ in the interval $0\leq t\leq T$, for specified duration $T$, which maximizes the final value $z(T)=X(T)$.

The first step towards finding the optimal solution is the formation of the so-called control Hamiltonian \cite{Bryson75} corresponding to system (\ref{xyz}), (\ref{theta}). The control Hamiltonian is a mathematical construction and its maximization leads to the maximization of the target quantity, in the present case $z(T)$. It is formulated by adjoining to each state equation a Lagrange multiplier as below
\begin{eqnarray}
\label{Hc}
H_c&=&\lambda_x\dot{x}+\lambda_y\dot{y}+\lambda_z\dot{z} +\mu u\nonumber\\
&=&(\lambda_x z-\lambda_z x+\mu)u+\frac{1}{2}\lambda_x y-\frac{\Gamma}{2}\lambda_y y-\frac{1}{2}\lambda_y x,
\end{eqnarray}
where $\lambda_x,\lambda_y,\lambda_z,\mu$ are the Lagrange multipliers corresponding to state variables $x,y,z,\theta$. These multipliers obey the adjoint set of equations
\begin{equation}
\label{adjoint}
\left(\begin{array}{c}
    \dot{\lambda}_z  \\
    \dot{\lambda}_y \\
    \dot{\lambda}_x
\end{array}\right)
=
\left(\begin{array}{c}
    -\frac{\partial H_c}{\partial z}  \\\noalign{\vskip3pt}
    -\frac{\partial H_c}{\partial y} \\\noalign{\vskip3pt}
    -\frac{\partial H_c}{\partial x}
\end{array}\right)
=
\left(\begin{array}{ccc}
    0 & 0 & -u(t) \\\noalign{\vskip3pt}
    0 & \frac{\Gamma}{2} & -\frac{1}{2} \\\noalign{\vskip3pt}
    u(t) & \frac{1}{2} & 0
  \end{array}\right)
\left(\begin{array}{c}
    \lambda_z  \\
    \lambda_y \\
    \lambda_x
\end{array}\right),
\end{equation}
while $\mu$ is constant because $\theta$ is a cyclic variable. For the problem at hand, where we seek to maximize $z(T)$, the multipliers should satisfy the following terminal conditions \cite{Bryson75}
\begin{eqnarray}
\lambda_x(T)&=0,\label{lx_T}\\
\lambda_y(T)&=0,\label{ly_T}\\
\lambda_z(T)&=1.\label{lz_T}
\end{eqnarray}
We also point out that, because of the construction of $H_c$, the state equations (\ref{xyz}) can also be formulated as $\dot{s}=\partial H_c/\partial \lambda_s$, $s=x, y, z$, justifying thus the term Hamiltonian.

According to optimal control theory \cite{Bryson75}, the control function $u(t)$ is chosen to maximize the control Hamiltonian $H_c$. In the case where the control is unbounded, even infinite values are allowed momentarily, which we call bang pulses, corresponding to instantaneous jumps in the angle $\theta$. Because $H_c$ is a linear function of $u$ with coefficient the \emph{switching} function
\begin{equation}
\label{switching_function}
\phi=\lambda_x z-\lambda_z x+\mu,
\end{equation}
if $\phi\neq 0$ for a finite time-interval then the corresponding optimal control should be $\pm\infty$ for the whole interval, which is clearly nonphysical. We deduce that in the case of unbounded $u$, the condition $\phi=0$ should hold almost everywhere, excluding some isolated points where jumps in the angle $\theta$ may take place. The optimal control preserving this condition is termed \emph{singular} \cite{Bryson75}. Singular controls frequently appear as parts of the optimal solution, see for example the famous aerospace Goddard problem \cite{Tsiotras92}, and have been also utilized in the context of dissipative two-level systems in order to minimize losses due to dissipation \cite{Lapert10,Lin20}. If there are bounds on $u(t)$, then the optimal pulse sequence may also contain finite time intervals where the control attains one of the boundary values, determined by the sign of the switching function $\phi$ so that $H_c$ is maximized.

\section{Optimal solution in the absence of dissipation, $\Gamma=0$}

\label{sec:sol0}

As a warm up example, we consider first the case without dissipation at the intermediate state, i.e. set $\Gamma=0$. In the absence of dissipation the vector $(x, y, z)^T$ describing the state of the system remains on the Bloch sphere, and a meaningful question is what is the minimum necessary time to return to the initial state at north pole while the angle $\theta$ changes from $0$ to $\pi/2$. We will not reformulate the minimum-time optimal control problem but rather use the formulation of the previous section in order to find the minimum-time solution. Note that this solution was obtained in Ref. \cite{Boscain02} in the original reference frame $XYZ$, as the minimum-time solution to bring the Bloch vector from the north pole $(X, Y, Z)^T=(0, 0, 1)$ to $(X, Y, Z)^T=(1, 0, 0)$ on the equator.

We start without putting any restrictions on the control $u(t)$, thus the optimal pulse sequence may consist only of bang pulses, which rotate instantaneously angle $\theta$ by a finite amount, separated by finite time intervals where the switching function (\ref{switching_function}) is zero and the control is singular. In order to find the singular control, we use condition $\phi=0$ as well as the additional relations $\dot{\phi}=\ddot{\phi}=0$, and obtain the equations
\begin{eqnarray}
\phi=0&\Rightarrow&z\lambda_x-x\lambda_z+\mu=0,\label{phi}\\
\dot{\phi}=0&\Rightarrow&z\lambda_y-y\lambda_z=0,\label{dphi}\\
\ddot{\phi}=0&\Rightarrow& \left(-\frac{1}{2}z+uy\right)\lambda_x-ux\lambda_y+\frac{1}{2}x\lambda_z=0.\label{ddphi}
\end{eqnarray}
Solving Eqs. (\ref{phi}), (\ref{dphi}) with respect to $\lambda_x, \lambda_y$ we find
\begin{eqnarray}
\bar{\lambda}_x&=&\frac{x}{z}\bar{\lambda}_z-\frac{1}{z},\label{lx}\\
\bar{\lambda}_y&=&\frac{y}{z}\bar{\lambda}_z,\label{ly}
\end{eqnarray}
where (recall that $\mu\neq 0$)
\begin{equation}
\label{lambda_i}
\bar{\lambda}_i=\frac{\lambda_i}{\mu},\quad i=x,y,z.
\end{equation}
If we plug Eqs. (\ref{lx}), (\ref{ly}) in Eq. (\ref{ddphi}), we get the singular feedback control
\begin{equation}
\label{singular_feedback}
u_s=\frac{z}{2y}.
\end{equation}
The control Hamiltonian is constant according to Maximum Principle \cite{Bryson75}, and on the singular arc this constancy condition gives the additional relation
\begin{equation}
\label{H_const_0}
H_c=c\Rightarrow\frac{1}{2}\lambda_x y-\frac{1}{2}\lambda_y x=c.
\end{equation}
If we plug Eqs. (\ref{lx}), (\ref{ly}) in Eq. (\ref{H_const_0}), we end up with the relation
\begin{equation}
\frac{y}{z}=-\frac{2c}{\mu},
\end{equation}
which implies that the singular control (\ref{singular_feedback}) is constant
\begin{equation}
\label{singular_control_0}
u_s=-\frac{\mu}{4c}.
\end{equation}
The singular arc lies on the singular surface
\begin{equation}
\label{singular_surface_0}
z-2u_sy=0,
\end{equation}
which is a plane parallel to $x$-axis passing through the origin.

\begin{figure}[!t]
 \centering
		\begin{tabular}{cc}
     	\subfigure[$\ $]{
	            \label{fig:zero_control}
	            \includegraphics[width=.5\linewidth]{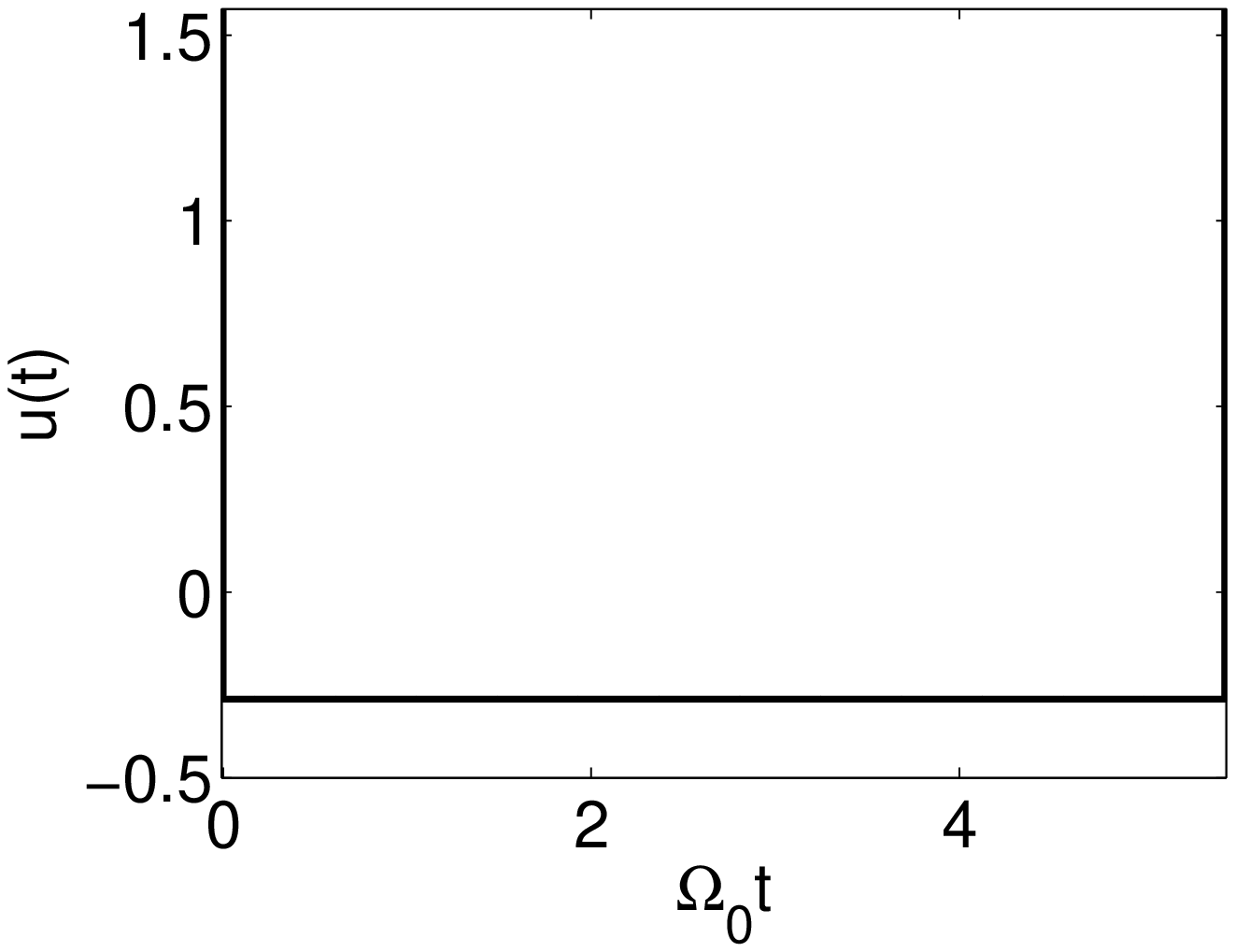}} &
       \subfigure[$\ $]{
	            \label{fig:zero_theta}
	            \includegraphics[width=.5\linewidth]{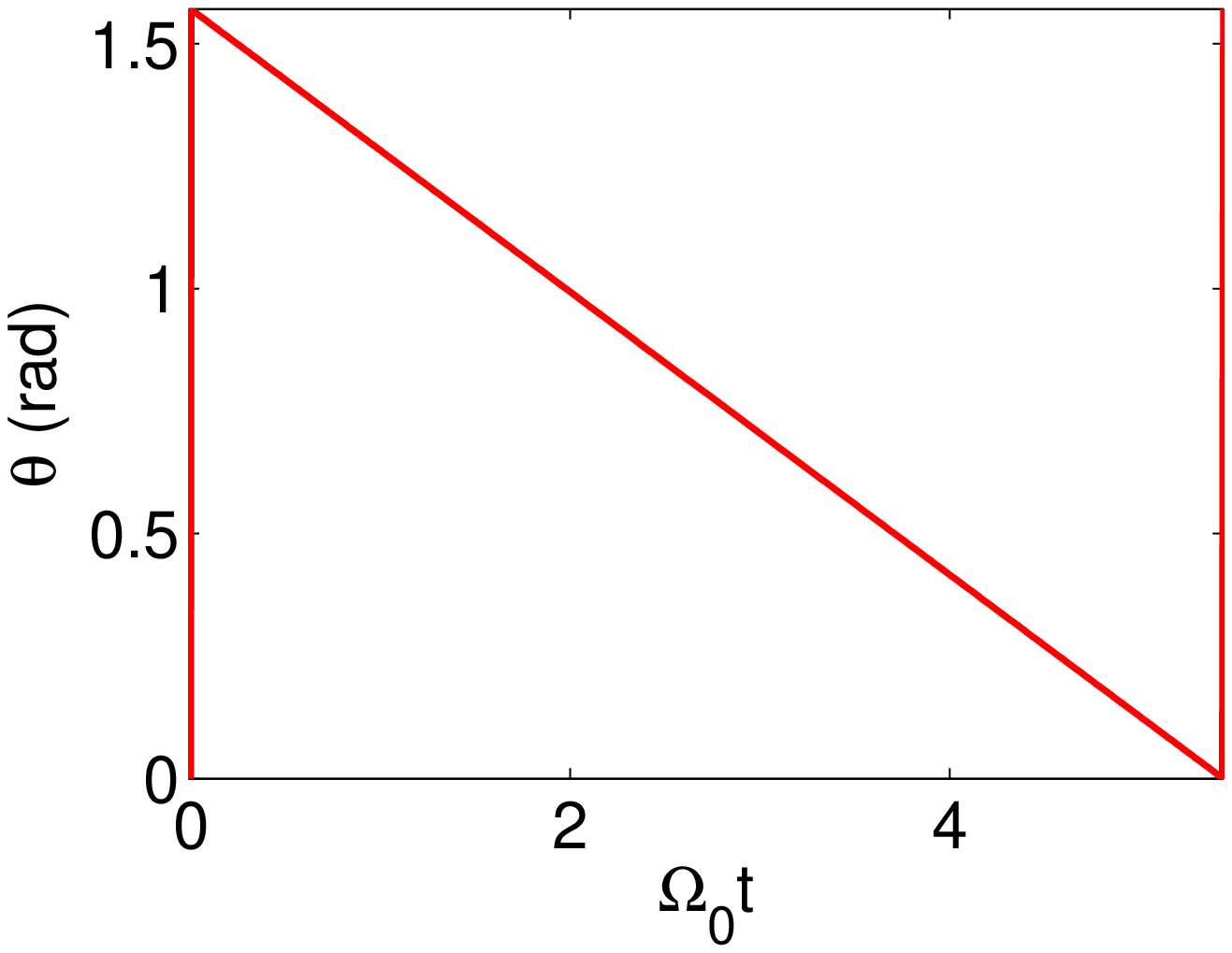}} \\
       \subfigure[$\ $]{
	            \label{fig:zero_populations}
	            \includegraphics[width=.5\linewidth]{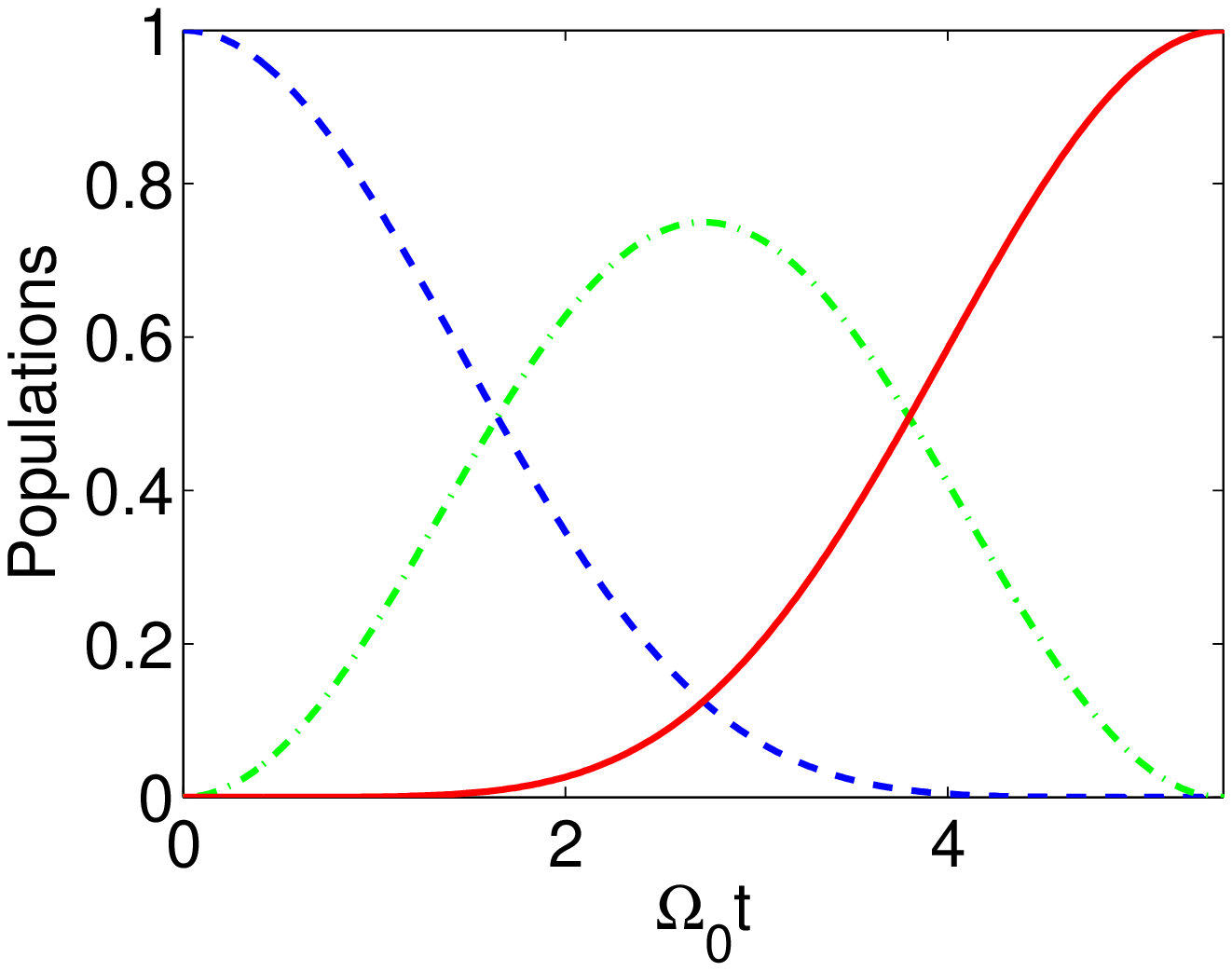}} &
       \subfigure[$\ $]{
	            \label{fig:zero_Bloch}
	            \includegraphics[width=.5\linewidth]{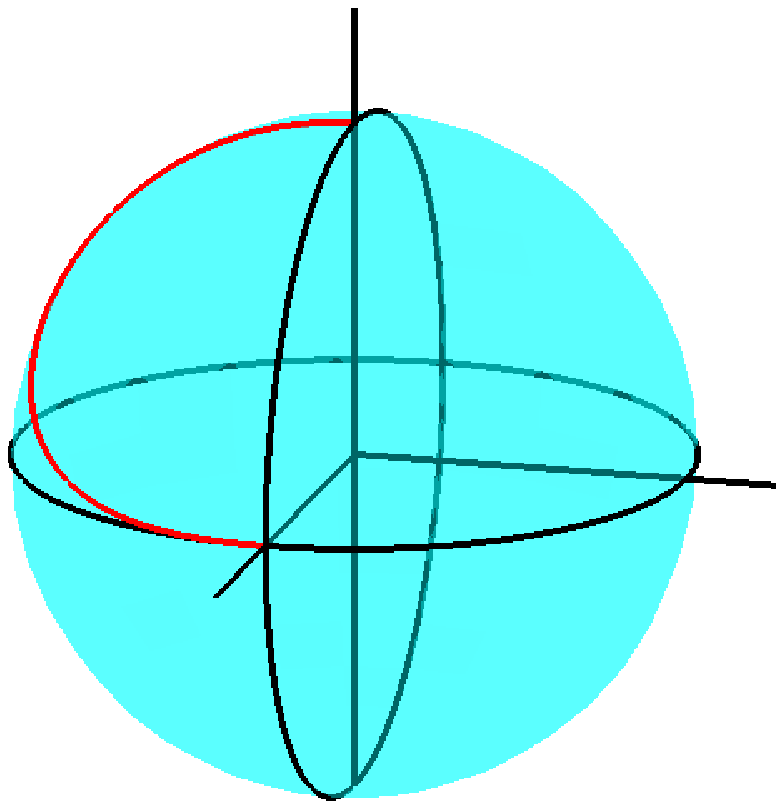}}
		\end{tabular}
\caption{(a) Minimum-time optimal control $u=\dot{\theta}$ for complete population transfer from state $|1\rangle$ to state $|3\rangle$ in the absence of dissipation ($\Gamma=0$). (b) Evolution of the polar angle $\theta$ of the total field. (c) Populations of the initial state $|1\rangle$ (blue dashed line), intermediate state $|2\rangle$ (green dashed-dotted line), and target state $|3\rangle$ (red solid line). (d) Optimal trajectory (red solid line) on the Bloch sphere, in the original reference frame $XYZ$.}
\label{fig:zero}
\end{figure}

Having found the singular control and the singular surface, we are now in a position to describe the optimal pulse sequence, which has the bang-singular-bang form. The only way to reach the singular surface with a bang pulse when starting from the north pole in the transformed frame is with an initial $\pi/2$ bang pulse, which rotates instantaneously the Bloch vector from the starting point $(0, 0, 1)^T$ to $(1, 0, 0)^T$. Then, a gradual rotation of duration $T$ under the constant singular control $u_s$ brings the Bloch vector to $(-1, 0, 0)^T$, and finally another $\pi/2$ bang pulse returns the Bloch vector back to $(0, 0, 1)^T$. The value of the constant singular control $u_s$ is obtained from the requirement that after the second bang pulse it should be $\theta(T^+)=\pi/2$. But
\begin{equation}
\label{change_theta}
\theta(T^+)=\frac{\pi}{2}+\theta(T^-)=\frac{\pi}{2}+u_sT+\theta(0^+)=\pi+u_sT,
\end{equation}
thus
\begin{equation}
\label{min_sinhular}
u_s=-\frac{\pi}{2T}.
\end{equation}
During the singular part of the sequence the Bloch vector traces a half circle on the singular plane, rotated around the total field $(0, u_s, -1/2)^T$ with angular velocity
\begin{equation}
\omega=\sqrt{u_s^2+\left(\frac{1}{2}\right)^2}=\frac{1}{2}\sqrt{\frac{\pi^2}{T^2}+1}.
\end{equation}
The necessary duration $T$ to cover the half circle is obtained from the relation for the rotation angle $\varphi=\omega T=\pi$ and is found to be
\begin{equation}
\label{min_T}
T=\pi\sqrt{3}.
\end{equation}
We have just recovered the minimum-time solution obtained in Ref. \cite{Boscain02}. Note that the minimum time obtained there is actually $\pi\sqrt{3}/2$, since the authors use system (\ref{Schrodinger}) without the factor $1/2$ in the right hand side. This minimum-time solution actually corresponds to the intuitive pulse sequence, where the pump pulse precedes the Stokes pulse, as we immediately explain. In the original reference frame $XYZ$, the total field is initially aligned with the Bloch vector at the north pole, and the application of the first $\pi/2$ pulse brings it on the equator. Then, it is gradually rotated back to the north pole while the Bloch vector is rotated towards the equator. The final $\pi/2$ pulse brings the total field vector back to the equator, where is aligned again with the rotated Bloch vector. Between the two bang pulses the angle $\theta$ changes linearly with time
\begin{equation}
\theta=\frac{\pi}{2}+u_st=\frac{\pi}{2}\left(1-\frac{t}{T}\right),
\end{equation}
thus
\begin{eqnarray}
\Omega_p(t)&=&\sin\theta=\cos\left(\frac{\pi t}{2T}\right),\\
\Omega_s(t)&=&\cos\theta=\sin\left(\frac{\pi t}{2T}\right).
\end{eqnarray}
In Fig. \ref{fig:zero_control} we display the optimal control $u(t)$, with the initial and final bang pulses, and in Fig. \ref{fig:zero_theta} the corresponding evolution of angle $\theta$. In Fig. \ref{fig:zero_populations} we plot the populations of levels $|i\rangle$, $i=1,2,3$, and in Fig. \ref{fig:zero_Bloch} the optimal trajectory on the Bloch sphere in the original reference frame $XYZ$.

Although typically this minimum time solution can be considered as a shortcut, the fact that the angle $\theta$ changes back and forth between $0$ and $\pi/2$ makes it not very attractive. In order to obtain solution resembling the counterintuitive STIRAP pulse sequence, where the Stokes pulse precedes the pump pulse, it is necessary to impose restrictions on the control $u=\dot{\theta}$. But before doing so, it is instructive to consider first a pulse sequence inspired from the minimum-time solution, of the bang-constant-bang form, where the initial and final bang pulses have equal magnitude $\theta_0$ in the range $0\leq\theta_0\leq \pi/2$ while the control $u$ in the intermediate interval is constant.
The final condition $\theta(T)=\pi/2$ imposes the following constraint between $\theta_0, u, T$
\begin{equation}
\label{theta_condition}
2\theta_0+uT=\frac{\pi}{2}.
\end{equation}
Another condition is obtained from the requirement that the Bloch vector in the transformed $xyz$ frame should return to the north pole. In order to derive it, it is more convenient to work using the two-level system obtained from Eq. (\ref{xyz}) using the transformation
\begin{eqnarray}
x&=&b_1b_2^*+b_1^*b_2,\\
y&=&i(b_1b_2^*-b_1^*b_2),\\
z&=&|b_1|^2-|b_2|^2,
\end{eqnarray}
where the probability amplitudes $\mathbf{b}=(b_1, b_2)^T$ obey the Schr\"{o}dinger equation
\begin{equation}
\label{two-level}
i\dot{\mathbf{b}}=\frac{1}{2}\left[u(t)\sigma_y-\frac{1}{2}\sigma_z\right]\mathbf{b},
\end{equation}
with $\sigma_i, i=x,y,z$ being the Pauli matrices.
The two-level system propagator corresponding to the pulse sequence with initial and final bang pulses of magnitude $\theta_0$ and intermediate constant control $u$ for duration $T$ is
\begin{equation}
\label{propagator_a}
U=e^{-\frac{i}{2}\theta_0\sigma_y}e^{-\frac{i}{2}\varphi(n_y\sigma_y+n_z\sigma_z)}e^{-\frac{i}{2}\theta_0\sigma_y},
\end{equation}
where
\begin{equation}
\label{vphi}
\varphi=\omega T=T\sqrt{u^2+\frac{1}{4}}
\end{equation}
is the rotation angle around the total field $(0, u, -1/2)^T=\omega(0, n_y, n_z)^T$ and
\begin{eqnarray}
n_y&=&\frac{u}{\omega}=\frac{u}{\sqrt{u^2+\frac{1}{4}}},\label{ny}\\
n_z&=&-\frac{1}{2\omega}=-\frac{1}{2\sqrt{u^2+\frac{1}{4}}}.\label{nz}
\end{eqnarray}
Using the identity \cite{Merzbacher98}
\begin{equation}
\label{identity}
e^{-\frac{i}{2}\varphi\hat{\mathbf{n}}\cdot\hat{\mathbf{\sigma}}}=I\cos\frac{\varphi}{2}-i\hat{\mathbf{n}}\cdot\hat{\mathbf{\sigma}}\sin\frac{\varphi}{2}.
\end{equation}
we can find that
\begin{equation}
\label{propagator_b}
U=\left(\cos\theta_0\cos\frac{\varphi}{2}-n_y\sin\theta_0\sin\frac{\varphi}{2}\right)I-i\left(\sin\theta_0\cos\frac{\varphi}{2}+n_y\cos\theta_0\sin\frac{\varphi}{2}\right)\sigma_y-in_z\sin\frac{\varphi}{2}\sigma_z
\end{equation}
\begin{figure}[!t]
 \centering
		\begin{tabular}{cc}
     	\subfigure[$\ $]{
	            \label{fig:pos_zero_control}
	            \includegraphics[width=.5\linewidth]{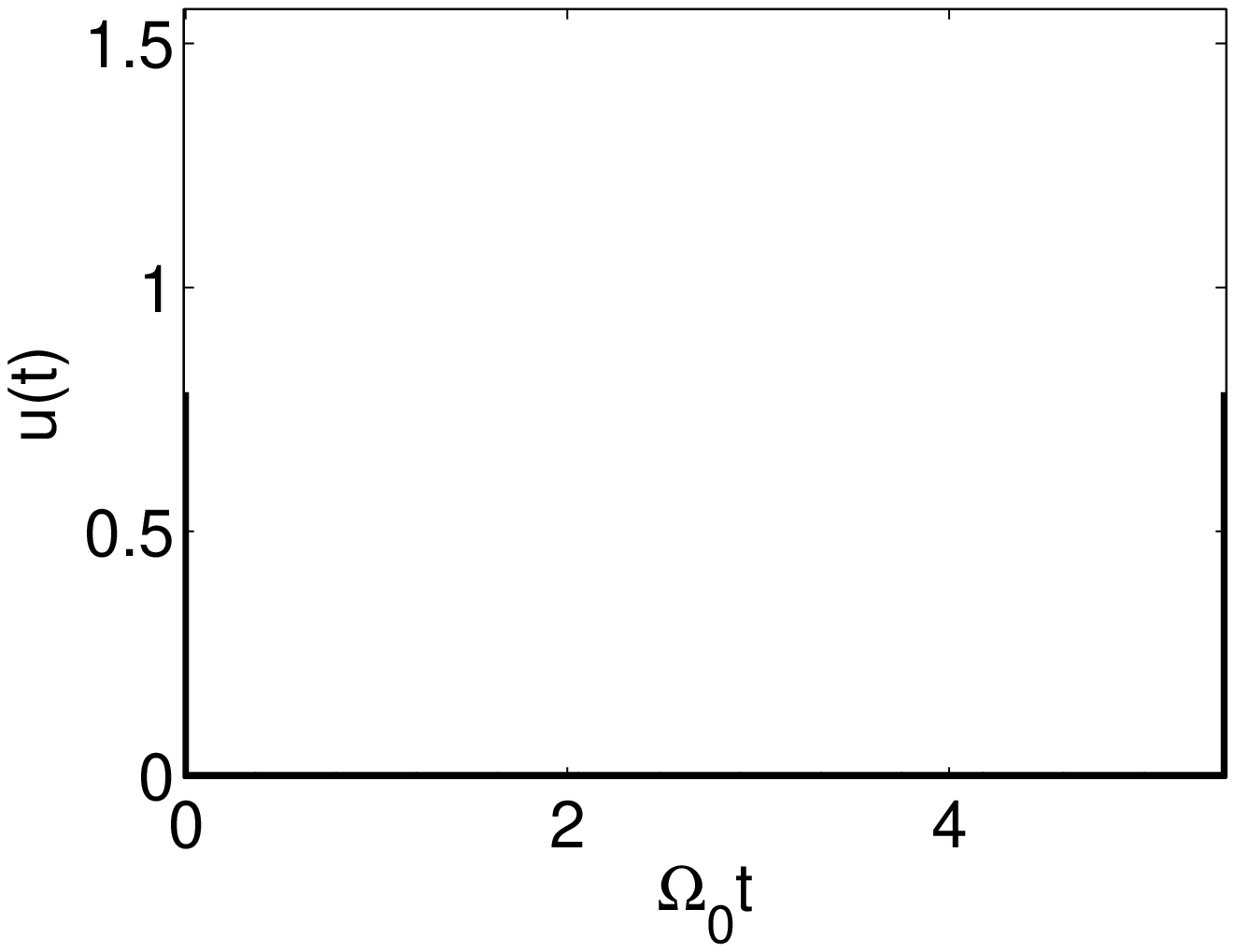}} &
       \subfigure[$\ $]{
	            \label{fig:pos_zero_theta}
	            \includegraphics[width=.5\linewidth]{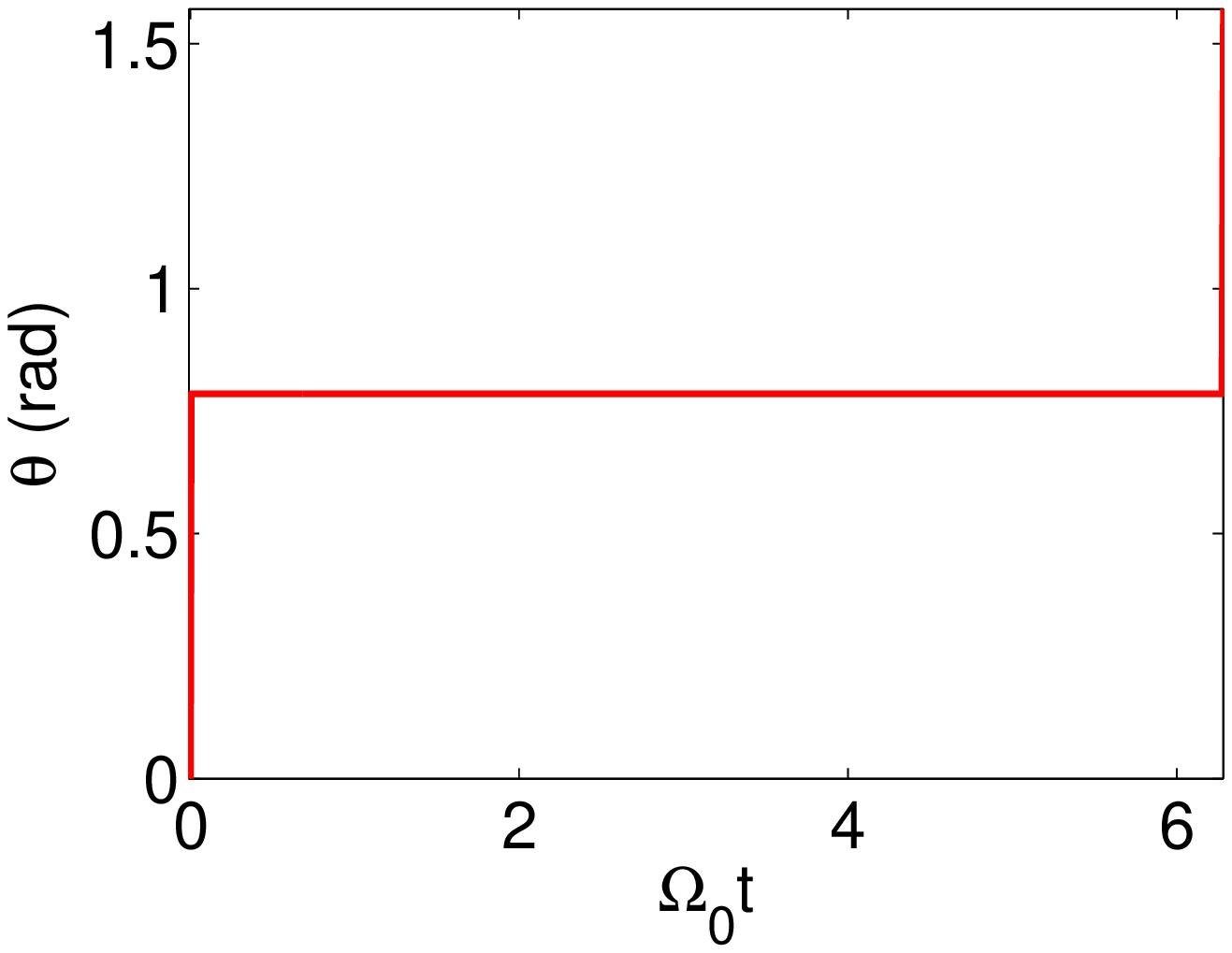}} \\
       \subfigure[$\ $]{
	            \label{fig:pos_zero_populations}
	            \includegraphics[width=.5\linewidth]{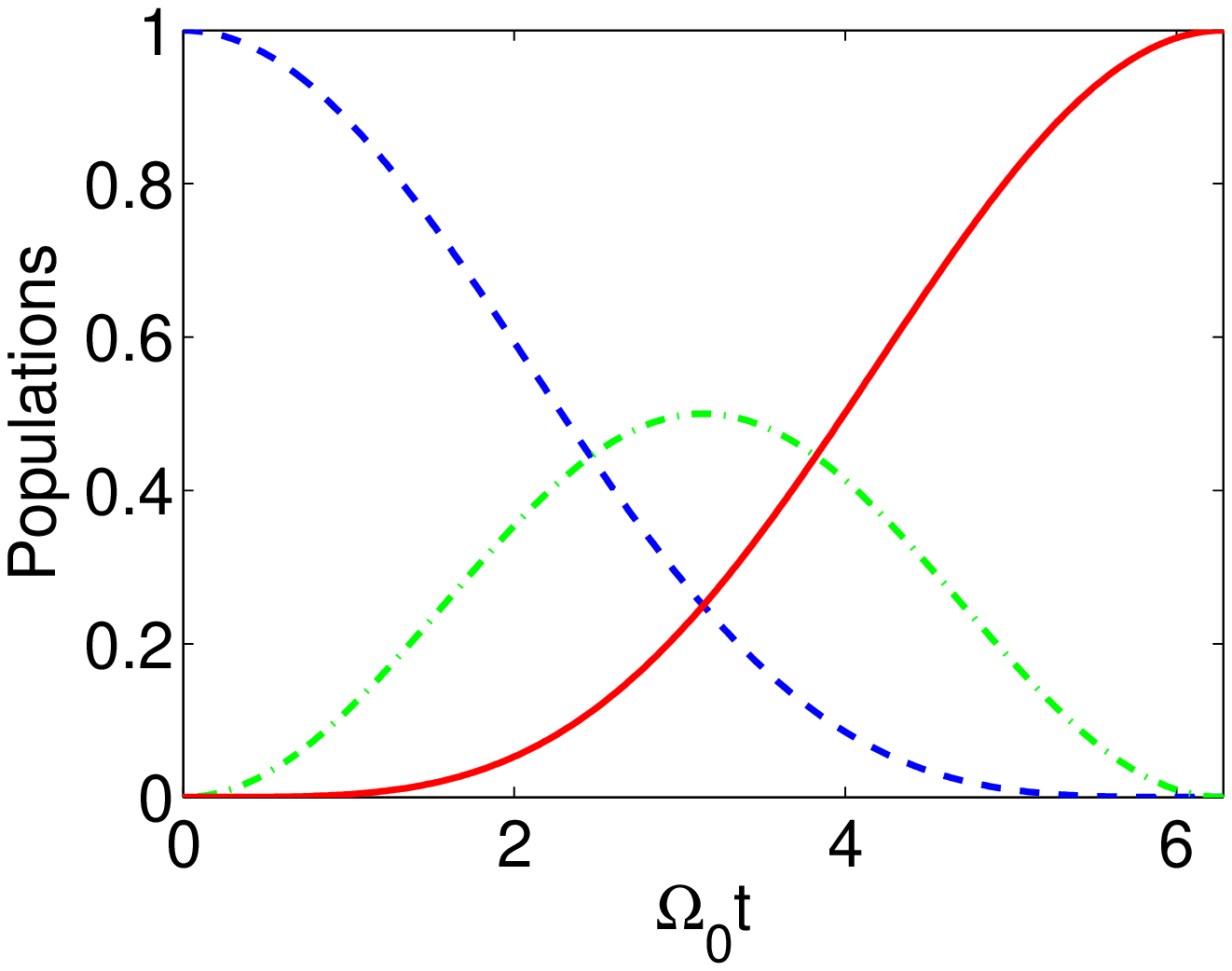}} &
       \subfigure[$\ $]{
	            \label{fig:pos_zero_Bloch}
	            \includegraphics[width=.5\linewidth]{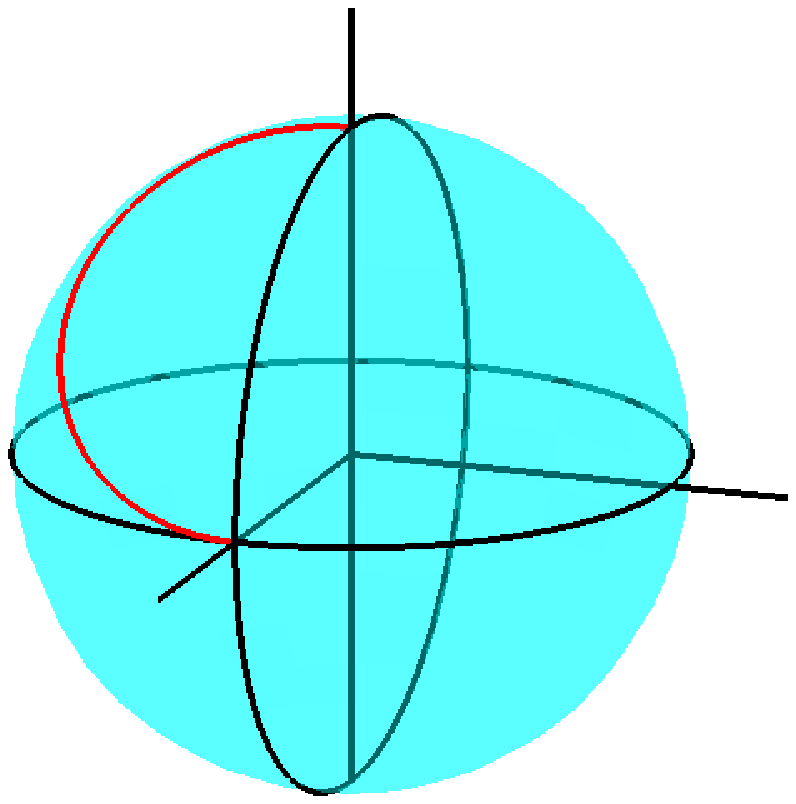}}
		\end{tabular}
\caption{(a) Nonnegative minimum-time optimal control $u=\dot{\theta}\geq 0$ for complete population transfer from state $|1\rangle$ to state $|3\rangle$ in the absence of dissipation ($\Gamma=0$). (b) Evolution of the polar angle $\theta$ of the total field. (c) Populations of the initial state $|1\rangle$ (blue dashed line), intermediate state $|2\rangle$ (green dashed-dotted line), and target state $|3\rangle$ (red solid line). (d) Optimal trajectory (red solid line) on the Bloch sphere, in the original reference frame $XYZ$.}
\label{fig:pos_zero}
\end{figure}
Note that the absence of a term proportional to $\sigma_x$ from the total propagator is due to the choice of equal bang pulses at the beginning and end. For the Bloch vector to return at the north pole at the final time, it is necessary to be zero the coefficient of $\sigma_y$ in the total propagator, thus
\begin{equation}
\label{condition}
\cot\frac{\varphi}{2}=-n_y\cot\theta_0.
\end{equation}

This is a second condition involving parameters $\theta_0, u, T$. Note that using condition (\ref{theta_condition}) we can eliminate $\theta_0$ from Eq. (\ref{condition}) and obtain the duration $T$ as an implicit function of $u$. Although the function $T(u)$ cannot be obtained in closed form, in the appendix we calculate the derivative
\begin{equation}
\label{dTdu}
\frac{dT}{du}=\frac{2\cot\theta_0}{u^2+\frac{1}{4}}\left(1-\frac{\varphi}{2}\cot\frac{\varphi}{2}\right)\geq 0
\end{equation}
and prove that $dT/du\geq 0$, i.e. $T$ is an increasing function of $u$. But from condition (\ref{theta_condition}) and for $0\leq\theta_0\leq \pi/2$ we obtain that
\begin{equation}
-\frac{\pi}{2T}\leq u\leq\frac{\pi}{2T}.
\end{equation}
The minimum duration is obtained at the lower bound $u=u_s=-\pi/(2T)$, corresponding to the minimum-time intuitive solution derived above, with $\theta_0=\pi/2$, $\varphi=\pi$ and $T=\pi\sqrt{3}$. If, on the other hand, we impose the condition
\begin{equation}
\label{positive_u}
u=\dot{\theta}\geq 0
\end{equation}
in order to avoid the back and forth changes in $\theta$, we find that the duration is minimized for $u=0$, thus the optimal pulse sequence has the bang-off-bang form with $\theta_0=\pi/4$, $\varphi=\pi$ and duration $T=2\pi$. The increased duration compared to the minimum value (\ref{min_T}) is the price paid for having the restriction (\ref{positive_u}) satisfied. Note that since in this case $u=0$ is the lower control bound, the bang-off-bang solution is actually of the bang-bang form. In Fig. \ref{fig:pos_zero_control} we plot the optimal control $u(t)$ and in Fig. \ref{fig:pos_zero_theta} the angle $\theta(t)$. In Fig. \ref{fig:pos_zero_populations} we display the populations of the three energy levels, while in Fig. \ref{fig:pos_zero_Bloch} the optimal trajectory in the original reference frame $XYZ$. Comparing Figs. \ref{fig:zero_populations}, \ref{fig:pos_zero_populations}, we observe that in the latter case, where the duration is longer, the intermediate state is less populated during the transfer.

\section{Optimal solution in the presence of dissipation, $\Gamma\neq 0$}

\label{sec:solG}

After setting the stage, we now move to the case with nonzero dissipation. We also impose the condition (\ref{positive_u}), in order to derive solutions with increasing $\theta$, conforming to the spirit of shortcuts to adiabaticity. Note that this restriction allows the possibility of time intervals where the control attains this lower bound, as we saw in the previous section. But before discussing the structure of the optimal pulse sequence, let us first derive how the singular control and the singular surface are modified in the presence of dissipation.

Eqs (\ref{phi}) and (\ref{dphi}) remain valid, while Eq. (\ref{ddphi}) is modified as follows
\begin{equation}
\label{G_ddphi}
\ddot{\phi}=0\Rightarrow \left(-\frac{1}{2}z+uy\right)\lambda_x+\left(\frac{\Gamma}{2}z-ux\right)\lambda_y+\left(\frac{\Gamma}{2}y+\frac{1}{2}x\right)\lambda_z=0.
\end{equation}
Solving for $\bar{\lambda}_i=\lambda_i/\mu$, $i=x,y,z$, we get
\begin{eqnarray}
\bar{\lambda}_x&=&\frac{x}{\Gamma yz^2}\left(uy-\frac{1}{2}z\right)-\frac{1}{z},\label{l1}\\
\bar{\lambda}_y&=&\frac{1}{\Gamma z^2}\left(uy-\frac{1}{2}z\right),\label{l2}\\
\bar{\lambda}_z&=&\frac{1}{\Gamma yz}\left(uy-\frac{1}{2}z\right).\label{l3}
\end{eqnarray}
On the singular arc, where $\phi=0$, the constant Hamiltonian becomes
\begin{equation}
\label{H_const}
H_c=c\Rightarrow\frac{1}{2}\lambda_x y-\frac{\Gamma}{2}\lambda_y y-\frac{1}{2}\lambda_y x=c.
\end{equation}
If we plug Eqs. (\ref{l1}), (\ref{l2}), (\ref{l3}) in Eq. (\ref{H_const}) we find the singular feedback control
\begin{equation}
\label{singular_control}
u_s=\frac{z}{y}\left(c_1\frac{z}{y}-\frac{1}{2}\right),
\end{equation}
where
\begin{equation}
\label{c1}
c_1=-\frac{2c}{\mu}.
\end{equation}
In order to find the equation of the singular surface, we will employ another constant of the motion of the system. Using Eqs. (\ref{xyz}) and (\ref{adjoint}), one can verify the following constant of the motion
\begin{equation}
\label{c_motion}
\bar{\lambda}_xx+\bar{\lambda}_yy+\bar{\lambda}_zz=\frac{c'}{\mu}=c_2,
\end{equation}
where we have expressed it directly in terms of $\bar{\lambda}_i$.
If we replace Eqs. (\ref{lx}), (\ref{ly}) in Eq. (\ref{c_motion}) we get
\begin{equation}
\label{lz}
\bar{\lambda}_z=\frac{1}{r^2}(c_2z+x), \quad r^2=x^2+y^2+z^2.
\end{equation}
If we plug Eqs. (\ref{lx}), (\ref{ly}), (\ref{lz}) in Eqs. (\ref{H_const}), we find the equation of the singular surface
\begin{equation}
\label{singular_surface}
\frac{y}{z}+\Gamma\left(\frac{y}{r}\right)^2\left (c_2+\frac{x}{z}\right)=c_1.
\end{equation}
Note that for $\Gamma=0$ the singular surface reduces to the plane (\ref{singular_surface_0}).

We now move to discuss the structure of the optimal pulse sequence. We remind that the allowed values of the control are infinite bangs, which result in finite positive changes of the angle $\theta$, the lower bound value $u=0$, and the singular control (\ref{singular_control}) on the singular surface. Starting from $(x, y, z)^T=(0, 0, 1)$ the value $u=0$ is rejected because no evolution takes place in this case. We next ask the question whether the initial point can belong to the singular surface. Setting $y=0$ in Eq. (\ref{singular_surface}) we obtain $c_1=0$ and thus from Eq. (\ref{c1}) $c=0$. This means that the control Hamiltonian should be zero throughout, which corresponds to the free time solution \cite{Bryson75}. It actually corresponds to the infinite time solution where state $|2\rangle$ is not populated and the population is eventually transferred to state $|3\rangle$ without losses. In practice, state $|2\rangle$ is slightly populated and the singular control is used to maintain this small amount. But for finite durations $T$ that we consider here this case is not encountered. Thus the optimal pulse sequence starts with a bang control which changes angle $\theta$ from $\theta(0^-)=0$ to $\theta(0^+)=\theta_0$. We next show that the initial bang pulse cannot be followed by a singular control. From Eq. (\ref{xyz}) we see that the bang pulse also changes $x, z$ instantaneously, while it does not change $y$, which maintains its initial value $y=0$. For $z(0^+)\neq 0$, we find using similar arguments as before that only the infinite-time adiabatic control can serve as a singular control, which is rejected because of the finite available duration. For $z(0^+)=0$ we see from Eq. (\ref{singular_control}) that $c_1$ can have a nonzero value, thus a singular control seems to be possible. But in order to bring $z(0^+)=0$ from the starting point, it is necessary to apply a $\theta_0=\pi/2$ pulse. After that and due to the restriction (\ref{positive_u}) we can only apply $u=0$ for the rest of the time, so $z(T)=z(0^+)=0$ and obviously this protocol is not optimal. Note that if there is no restriction on the sign of the control, we simply recover the bang-singular-bang solution of the previous section, but in the presence of this condition the initial bang pulse can only be followed by an off pulse. Next, we examine whether the singular arc can be a terminal arc of the optimal trajectory. For this to be the case, it is necessary that condition (\ref{ly_T}) is satisfied. From Eq. (\ref{ly}), which holds on the singular arc, we see that $\lambda_y=0$ leads to $y=0$ or $\lambda_z=0$. The former choice is rejected for finite time solutions that we seek here, while the latter contradicts condition (\ref{lz_T}). Thus the optimal trajectory cannot terminate with a singular arc. We also show that the singular control cannot be followed by a terminal bang. From the adjoint system (\ref{adjoint}) we see that the value of $\lambda_y$ does not change by a bang control, thus it is necessary to satisfy condition (\ref{ly_T}) before leaving the singular arc. As we discussed previously, this leads to $y=0$, which is rejected for finite durations, or to $\lambda_z=0$. Since $\lambda_y=\lambda_z=0$ just before leaving the singular arc, a $\pi/2$ pulse is necessary in order to satisfy condition (\ref{lz_T}), which is forbidden under constraint (\ref{positive_u}). Of course, in the absence of the latter constraint this case is allowed and corresponds to the bang-singular-bang solution obtained in the previous section. But in the presence of the constraint an off pulse should follow the singular control, preceding the final bang pulse. We have thus derived that the form of the optimal pulse sequence under constraint (\ref{positive_u}) is bang-off-singular-off-bang, where note that the singular part appears for longer durations, as we discuss in the following section. Next, we show that the optimal pulse sequence is symmetric around the central time $t=T/2$.

Using system (\ref{adjoint}) we can find the equations in backward time $\sigma=T-t$ for $(\tilde{\lambda}_x(\sigma), \tilde{\lambda}_y(\sigma), \tilde{\lambda}_z(\sigma))=(-\lambda_x(T-\sigma), \lambda_y(T-\sigma), \lambda_z(T-\sigma))$ as follows
\begin{equation}
\label{b_adjoint}
\left(\begin{array}{c}
    \dot{\tilde{\lambda}}_z  \\
    \dot{\tilde{\lambda}}_y \\
    \dot{\tilde{\lambda}}_x
\end{array}\right)
=
\left(\begin{array}{ccc}
    0 & 0 & -\tilde{u}(\sigma) \\\noalign{\vskip3pt}
    0 & -\frac{\Gamma}{2} & -\frac{1}{2} \\\noalign{\vskip3pt}
    \tilde{u}(\sigma) & \frac{1}{2} & 0
  \end{array}\right)
\left(\begin{array}{c}
    \tilde{\lambda}_z  \\
    \tilde{\lambda}_y \\
    \tilde{\lambda}_x
\end{array}\right),
\end{equation}
with initial conditions $(\tilde{\lambda}_x(0), \tilde{\lambda}_y(0), \tilde{\lambda}_z(0))=(-\lambda_x(T), \lambda_y(T), \lambda_z(T))=(0, 0, 1)$ and $\tilde{u}(\sigma)=u(T-\sigma)$. Analogously, we can find the backward equations for $(\tilde{x}(\sigma), \tilde{y}(\sigma), \tilde{z}(\sigma))=(-x(T-\sigma), y(T-\sigma), z(T-\sigma))$
\begin{equation}
\label{b_xyz}
\left(\begin{array}{c}
    \dot{\tilde{z}}  \\
    \dot{\tilde{y}} \\
    \dot{\tilde{x}}
\end{array}\right)
=
\left(\begin{array}{ccc}
    0 & 0 & -\tilde{u}(\sigma) \\\noalign{\vskip3pt}
    0 & \frac{\Gamma}{2} & -\frac{1}{2} \\\noalign{\vskip3pt}
    \tilde{u}(\sigma) & \frac{1}{2} & 0
  \end{array}\right)
\left(\begin{array}{c}
    \tilde{z}  \\
    \tilde{y} \\
    \tilde{x}
\end{array}\right),
\end{equation}
with terminal conditions $(\tilde{x}(T), \tilde{y}(T), \tilde{z}(T))=(-x(0), y(0), z(0))=(0, 0, 1)$ and $\tilde{u}(\sigma)=u(T-\sigma)$. It can be easily verified that the backward equations (\ref{b_adjoint}), (\ref{b_xyz}) are adjoint to each other. Now observe that the backward equations for $(\tilde{\lambda_x}, \tilde{\lambda}_y, \tilde{\lambda}_z)$ and $(\tilde{x}, \tilde{y}, \tilde{z})$ are exactly the same as the forward equations for $(x, y, z)$ and
$(\lambda_x, \lambda_y, \lambda_z)$, respectively. Thus, if we treat $(\tilde{\lambda_x}, \tilde{\lambda}_y, \tilde{\lambda}_z)$ as the backward state variables and $(\tilde{x}, \tilde{y}, \tilde{z})$ as the backward adjoint variables, then the optimal control maximizing $\tilde{\lambda}_z(\sigma=T)$ is simply $\tilde{u}(\sigma)=u(\sigma)$, where $u(t)$ the optimal control in the forward direction. But $\tilde{\lambda}_z(\sigma=T)=\lambda_z(t=0)=z(t=T)$, where the last equality is obtained from the constant of the motion (\ref{c_motion}) evaluated at $t=0$ and $t=T$. Thus it is also $\tilde{u}(\sigma)=u(T-\sigma)$ and thus $u(\sigma)=u(T-\sigma)$ or
\begin{equation}
\label{control_symmetry}
u(t)=u(T-t),\quad 0\leq t\leq T.
\end{equation}
This means that the initial and final bang pulses have the same strength, and the off pulses have the same duration.
In the next paragraph we describe how to use this symmetry in order to numerically find the optimal pulse sequence for a given duration $T$.

Let $\theta_0$ be the rotation angle corresponding to the initial and final bang pulses, and $t_1$ the duration of the off pulses. After the initial bang and the first off pulses, the system arrives on the singular surface. The coordinates of the entry point are functions of $\theta_0, t_1$, while at the same time they satisfy Eq (\ref{singular_surface}) of the singular surface. We thus obtain a relation between $\theta_0, t_1$ and the constants $c_1, c_2$, which can typically be used to express $c_2$ in terms of $\theta_0, t_1, c_1$. Next, we apply the singular control (\ref{singular_control}) for duration $T-2t_1$. Since the singular control depends only on $c_1$, the coordinates of the exit point from the singular surface are functions of $\theta_0, t_1, c_1$. The constant $c_1$ can be expressed in terms of $\theta_0, t_1$ using the requirement $\theta=\pi/2-\theta_0$ at the exit point, which is obtained from the final value $\theta(T^+)=\pi/2$ and the fact that $\theta$ remains constant during the off pulses. Using the condition at the entry point, the constant $c_2$ can also be expressed in terms of $\theta_0, t_1$, as well as the coordinates of the exit point. From Eqs. (\ref{lx}), (\ref{ly}), (\ref{lz}), we see that the values of $\bar{\lambda}_i, i=x, y, z$ at the exit point can also be expressed as functions of $\theta_0, t_1$. The final step is, with these initial conditions at $t=T-t_1$, to propagate the adjoint system (\ref{adjoint}) (the corresponding one for $\bar{\lambda}_i$), for the final off pulse with duration $t_1$ and the final bang pulse of strength $\theta_0$. The optimal values of $\theta_0, t_1$ are obtained from the requirement to satisfy the two final conditions (\ref{lx_T}) and (\ref{ly_T}). Note that as long as the optimal $\theta_0, t_1$ are determined, the value $\bar{\lambda}_z(T)$ is also fixed, as expected from the constant of the motion (\ref{c_motion}). But from condition (\ref{lz_T}) we have $\bar{\lambda}_z(T)=1/\mu$, thus we can also find constant $\mu$. Using $\mu$ and the values of $c_1, c_2$, obtained for the optimal $\theta_0, t_1$, we can also find the initial constants $c, c'$.

\section{Example and discussion}

\label{sec:example}

\begin{figure}[!t]
 \centering
		\begin{tabular}{cc}
     	\subfigure[$\ $]{
	            \label{fig:G_control}
	            \includegraphics[width=.5\linewidth]{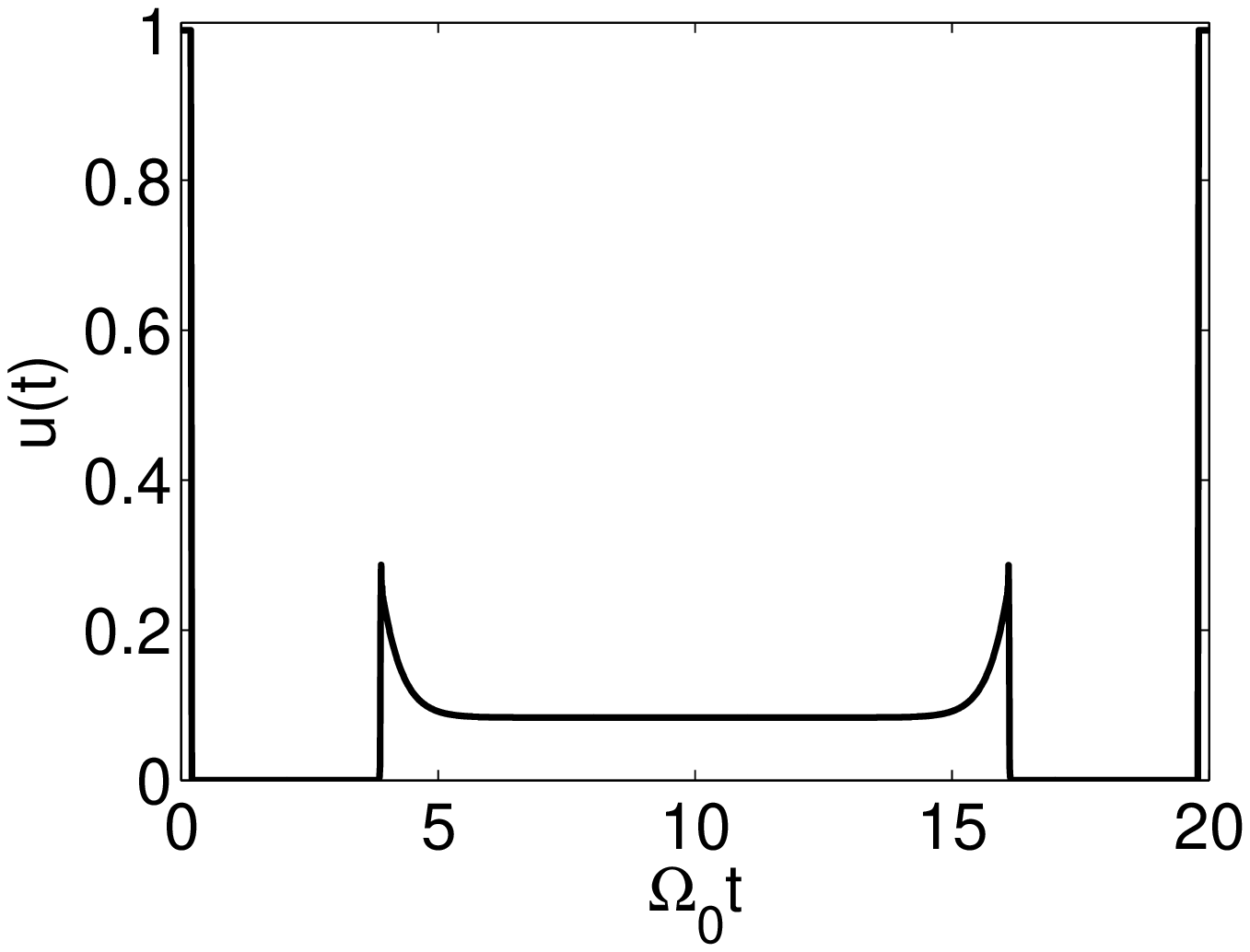}} &
       \subfigure[$\ $]{
	            \label{fig:G_theta}
	            \includegraphics[width=.5\linewidth]{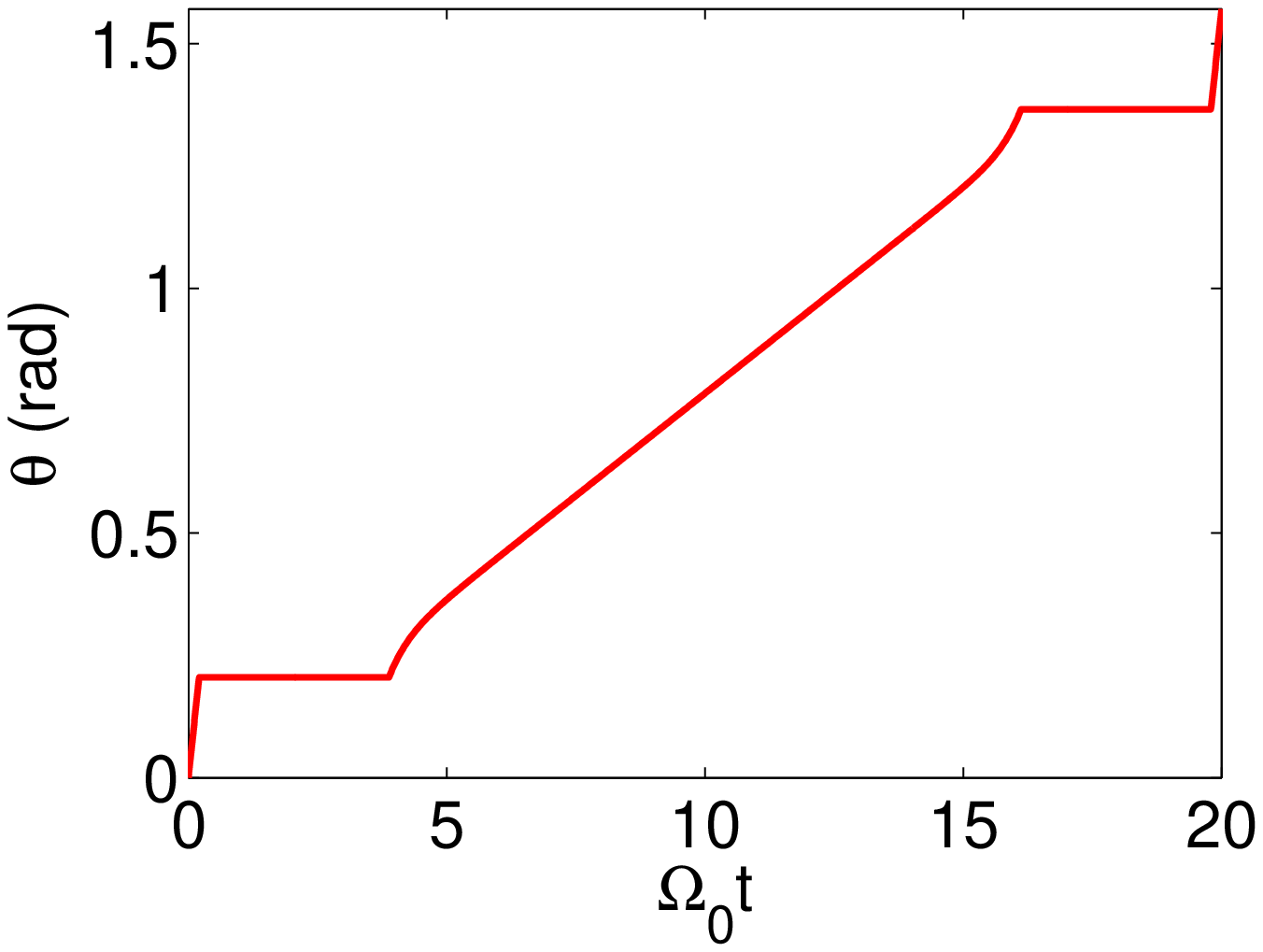}} \\
       \subfigure[$\ $]{
	            \label{fig:G_populations}
	            \includegraphics[width=.5\linewidth]{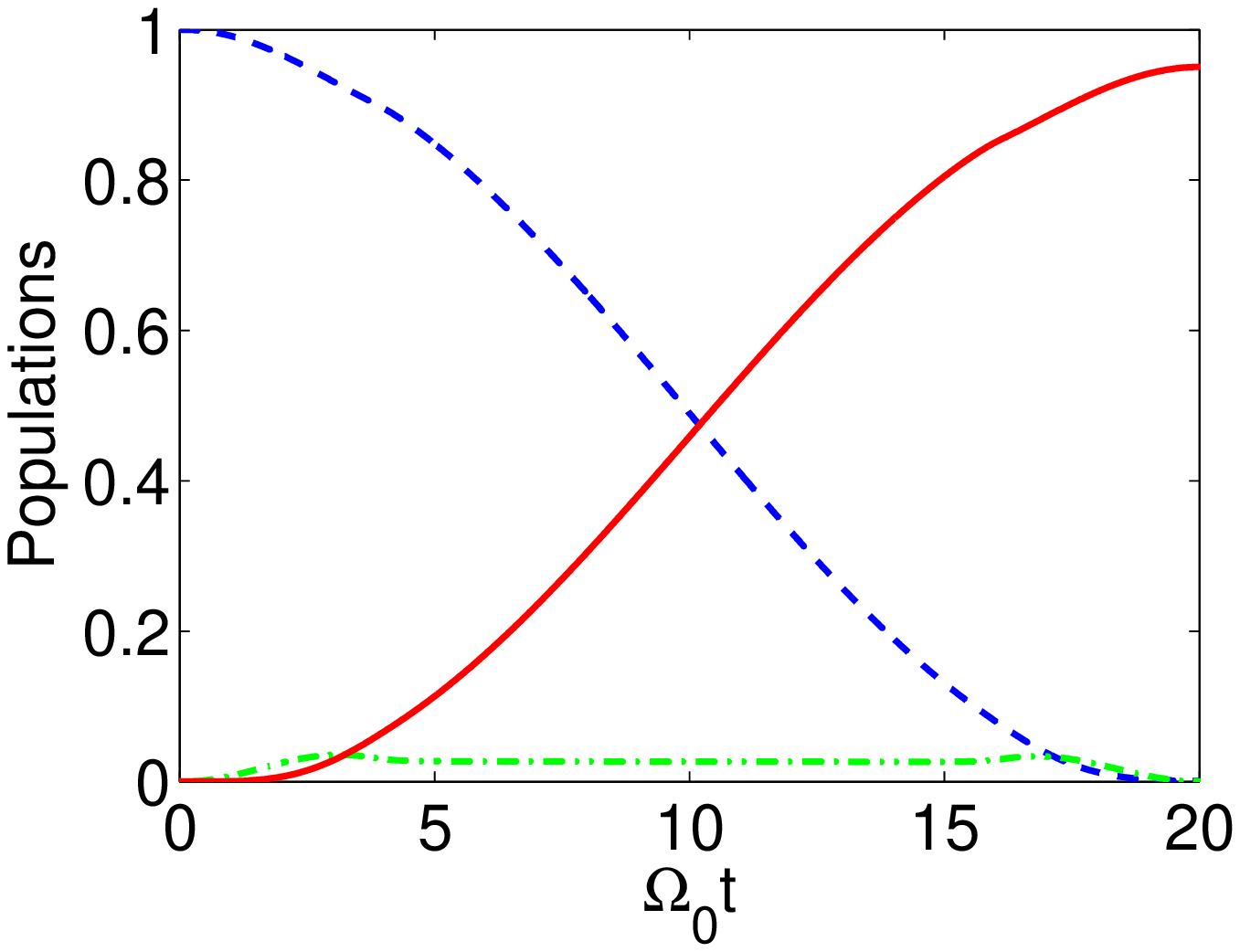}} &
       \subfigure[$\ $]{
	            \label{fig:G_Bloch}
	            \includegraphics[width=.5\linewidth]{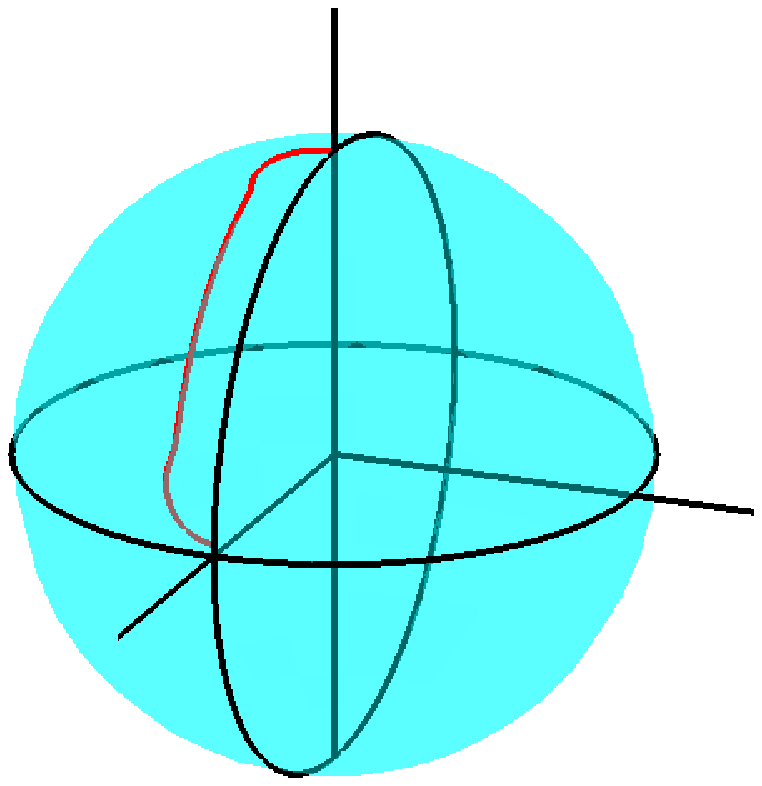}}
		\end{tabular}
\caption{(a) Optimal control $0\leq u(t)\leq 1$ maximizing the population of state $|3\rangle$ at the final time $T=20/\Omega_0$, in the presence of dissipation $\Gamma/\Omega_0=0.1$. (b) Evolution of the polar angle $\theta$ of the total field. (c) Populations of the initial state $|1\rangle$ (blue dashed line), intermediate state $|2\rangle$ (green dashed-dotted line), and target state $|3\rangle$ (red solid line). (d) Optimal trajectory (red solid line) in the original reference frame $XYZ$.}
\label{fig:G}
\end{figure}

In this section we consider a specific example with nonzero $\Gamma/\Omega_0=0.1$ and normalized duration $\Omega_0T=20$, and exploit the results obtained in order to discuss various aspects of the optimal solution. We take the control $u(t)$ to be bounded as follows
\begin{equation}
\label{control_bounds}
0\leq u(t)\leq 1,
\end{equation}
where the upper bound is imposed in order to avoid abrupt changes in the angle $\theta$, conforming thus more to the spirit of shortcuts to adiabaticity; as we will see, this bound does not affect much the optimal solution. In order to find the optimal solution we do not follow the procedure described in the last paragraph of the previous section but use instead the optimal control solver BOCOP \cite{bocop}. The numerically obtained optimal control is displayed in Fig. \ref{fig:G_control}, where observe that it is symmetric around the central time $t=10/\Omega_0$, as it was theoretically shown in the previous section. The symmetric initial and final bang pulses have finite height, due to the upper control bound imposed in (\ref{control_bounds}), while the off pulses have the same duration. The singular control occupies the central part of the pulse sequence and has a ``batman cowl" shape. The polar angle $\theta(t)$ of the total field is displayed in Fig. \ref{fig:G_theta}. Observe that, because of the finite height of the bang pulses, there are no abrupt changes in the angle $\theta$ at the beginning and end, but it rises following a ramp. During the off pulses it remains constant, while it increases in the singular control interval. The populations of the states are displayed in Fig. \ref{fig:G_populations}, where now the lossy intermediate state $|2\rangle$ is much less populated compared to the previous examples because of the larger available duration. This is also obvious in Fig. \ref{fig:G_Bloch}, where the optimal trajectory is displayed in the original reference frame $XYZ$. Note that due to the losses this trajectory does not actually lie on the Bloch sphere but goes slightly inside the Bloch ball.

Using Fig. \ref{fig:G_Bloch} we can explain the role of the bang and off pulses. Initially $\theta=0$ and the total field points to the north pole, aligned with the Bloch vector. The initial bang pulse immediately rotates the field to a finite $\theta>0$, and during the first off pulse the Bloch vector is rotated around this field such that a finite population in the intermediate state $|2\rangle$ is quickly built ($|C_2|^2=Y^2=y^2$), in order to accomplish the transfer from state $|1\rangle$ to state $|3\rangle$ in the finite available duration $T$. During the second off pulse the Bloch vector is rotated around the total field so the population $|C_2|^2=Y^2=y^2$ is quickly reduced, since the transfer approaches to the end and this population is not necessary. The final bang pulse brings the total field on the equator. Note that in the examples displayed in Figs. \ref{fig:zero}, \ref{fig:pos_zero} where, in the absence of dissipation, the complete population transfer is accomplished in shorter times, we have stronger bang pulses and the subsequent rotations of the Bloch vector lead to the buildup of a larger intermediate population.

\begin{figure}[!t]
\centering\includegraphics[width=0.6\linewidth]{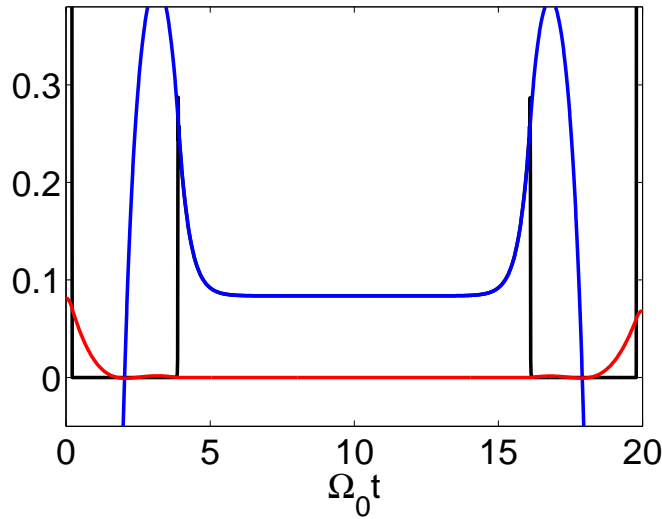}
\caption{Singular feedback control (blue solid line) from Eq. (\ref{singular_control}), fitting the numerically obtained optimal control (black solid line) for $c_1=-0.081368$, in the time interval of interest (between the ``ears" of ``batman cowl"). The red solid line shows the difference between the two sides of the singular surface equation (\ref{singular_surface}) for $c_1=-0.081368$ and $c_2=31.556$, which is zero in the interval of interest.}
\label{fig:G_singular}
\end{figure}

In order to verify that the singular control is indeed given by the feedback law (\ref{singular_control}), we use $y(t), z(t)$ obtained from the optimal solution and try to find a constant $c_1$ such that $u_s(t)$ from Eq. (\ref{singular_control}) fits the numerically found singular optimal control displayed in Fig. \ref{fig:G_control} (batman cowl). In Fig. \ref{fig:G_singular} we plot $u_s(t)$ for $c_1=-0.081368$ with blue solid line; observe the very good fit with the numerical optimal control in the interval of interest, between the ``ears" of the ``batmam cowl". Also, in order to verify that during this interval the optimal trajectory lies on the singular surface (\ref{singular_surface}), we use $x(t), y(t), z(t)$ obtained from the optimal solution and the value $c_1=-0.081368$ previously found, and seek a constant $c_2$ such that Eq. (\ref{singular_surface}) is satisfied. In Fig. \ref{fig:G_singular} we plot the difference between the two sides of Eg. (\ref{singular_surface}) for $c_2=31.556$ with red solid line. Observe that in the interval of interest this difference is zero.

\begin{figure}[!t]
\centering\includegraphics[width=0.6\linewidth]{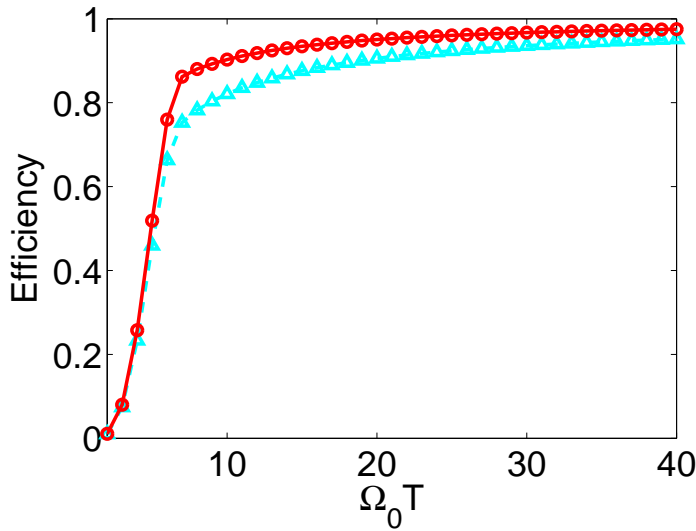}
\caption{Final population $|C_3(T)|^2$ of state $|3\rangle$ obtained with the optimal pulse sequence, as a function of duration $T$, for $\Gamma/\Omega_0=0.1$ (red circles) and $\Gamma/\Omega_0=0.2$ (cyan triangles).}
\label{fig:efficiency}
\end{figure}

In Fig. \ref{fig:efficiency} we plot the final population $|C_3(T)|^2$ of state $|3\rangle$, obtained from the solution of the optimal control problem, as a function of duration $T$, for $\Gamma/\Omega_0=0.1$ (red circles) and $\Gamma/\Omega_0=0.2$ (cyan triangles). Observe that, as the available duration increases, the population transfer also increases; the efficiency actually tends to unity for very large $T$. As expected, the rise of efficiency with increasing duration $T$ is higher for the case with lower dissipation (lower $\Gamma$). A quick numerical investigation reveals that for small values of dissipation the singular control appears in the optimal pulse sequence as the available duration $T$ approaches $2\pi$, the duration of the minimum-time solution for complete population transfer in the absence of dissipation and under the constraint $u\geq 0$. For larger dissipation values, singular control enters the optimal pulse sequence for shorter durations. Also, as the available duration increases, the singular arc occupies a larger portion of the optimal trajectory. We conclude that it is the singular control which provides paths in the Bloch ball where dissipation is minimized, leading thus to larger efficiencies for longer time intervals.

We close this section by pointing to two directions for possible future work. The first direction is to try to better integrate the current work within the framework of shortcuts to adiabaticity. For example, instead of using the ramp-constant and constant-ramp sequences in Fig. \ref{fig:G_theta} in order to connect the initial and final values of $\theta$ with the singular part, one may explore more smooth choices, as in Ref. \cite{Martikyan20a}. The second direction is to use the transformed reference frame $xyz$ and try to exploit the analogy between spins and springs for the design of optimal controls \cite{Li17,Martikyan20b}. As is well known, this analogy works well for small values of the polar angle of the Bloch vector \cite{Martikyan20b}, thus the transformed frame $xyz$, where $z(0)=1$ and the goal is to maximize $z(T)$, is perfectly suitable for its application. Actually, it is straightforward to obtain the spring system corresponding to the spin system (\ref{xyz}) by setting
\begin{equation}
\label{spring_control}
v(t)=u(t)z(t)=\dot{\theta}(t)z(t),
\end{equation}
leading to
\begin{equation}
\label{spring}
\left(\begin{array}{c}
    \dot{y} \\
    \dot{x}
\end{array}\right)
=
\left(\begin{array}{cc}
    -\frac{\Gamma}{2} & -\frac{1}{2} \\\noalign{\vskip3pt}
    \frac{1}{2} & 0
  \end{array}\right)
\left(\begin{array}{c}
    y \\
    x
\end{array}\right)
+v
\left(\begin{array}{c}
    0 \\
    1
\end{array}\right)
\end{equation}
This observation opens up many possibilities for the design of optimal controls which maximize population transfer from state $|1\rangle$ to state $|3\rangle$ through the lossy state $|2\rangle$. Also note that the ``transverse" relaxation $\Gamma$ pulls the Bloch vector towards the $z$-axis, making thus the spins to springs mapping more accurate.

\section{Conclusion}

\label{sec:conclusion}

We have applied optimal control theory in order to find shortcuts to adiabaticity optimizing population transfer in a three-level STIRAP system, for a specific time duration and in the presence of dissipation in the intermediate state. We fixed the sum of the intensities of the pump and Stokes lasers and utilized the mixing angle of the two fields as the only control parameter. We derived the optimal variation of this angle and disclosed the role of singular control in minimizing the effect of dissipation for longer available durations.

\begin{acknowledgments}
The authors would like to thank Ph.D. candidate Vasileios Evangelakos for coining the term ``batman cowl" to describe the shape of singular control.
\end{acknowledgments}

\appendix*

\section{Derivation of $dT/du$}

\label{sec:appendix}

We derive the expression (\ref{dTdu}) for the derivative $dT/du$ and also prove that $dT/du\geq 0$. By taking the derivative with respect to $u$ of both sides of Eq. (\ref{condition}), using Eqs. (\ref{theta_condition}), (\ref{vphi}), (\ref{ny}), and rearranging terms we obtain
\begin{equation*}
\sqrt{u^2+\frac{1}{4}}\left(\frac{1}{\sin^2\frac{\varphi}{2}}-\frac{n_y^2}{\sin^2\theta_0}\right)\frac{dT}{du}=\frac{\cot\theta_0}{2\left(u^2+\frac{1}{4}\right)^{3/2}}+\frac{uT}{\sqrt{u^2+\frac{1}{4}}}\left(\frac{1}{\sin^2\theta_0}-\frac{1}{\sin^2\frac{\varphi}{2}}\right)
\end{equation*}
Using the relations
\begin{eqnarray*}
\frac{1}{\sin^2\theta_0}&=&1+\cot^2\theta_0,\\
\frac{1}{\sin^2\frac{\varphi}{2}}&=&1+\cot^2\frac{\varphi}{2}=1+n_y^2\cot^2\theta_0,
\end{eqnarray*}
where the latter was obtained using additionally condition (\ref{condition}), we get
\begin{equation*}
\frac{1}{4\sqrt{u^2+\frac{1}{4}}}\frac{dT}{du}=\frac{\cot\theta_0}{2\left(u^2+\frac{1}{4}\right)^{3/2}}\left(1+\frac{uT}{2}\cot\theta_0\right).
\end{equation*}
Using in the right hand side the relations
\begin{eqnarray*}
uT&=&n_y\varphi,\\
\cot\theta_0&=&-\frac{1}{n_y}\cot\frac{\varphi}{2},
\end{eqnarray*}
obtained from Eqs. (\ref{vphi}) and (\ref{condition}), respectively, we finally arrive at the expression
\begin{equation*}
\frac{dT}{du}=\frac{2\cot\theta_0}{u^2+\frac{1}{4}}\left(1-\frac{\varphi}{2}\cot\frac{\varphi}{2}\right).
\end{equation*}

We next show that $dT/du\geq 0$. Consider the function
\begin{equation*}
f(x)=1-x\cot{x}
\end{equation*}
for $0\leq x<\pi$, since the rotation angle $\phi$ lies in the interval $0\leq \phi<2\pi$. It is
\begin{equation*}
f(x)\rightarrow 0\quad \mbox{for}\quad x\rightarrow 0,
\end{equation*}
while its derivative is
\begin{equation*}
f'(x)=\frac{x-\frac{1}{2}\sin{2x}}{\sin^2x}.
\end{equation*}
For the numerator of $f'$,
\begin{equation*}
g(x)=x-\frac{1}{2}\sin{2x},
\end{equation*}
it is
\begin{equation*}
g'(x)=1-\cos{2x}\geq 0.
\end{equation*}
Thus $g(x)\geq g(0)=0$, consequently $f'(x)\geq 0$ and $f(x)\geq f(0)=0$, thus $dT/du\geq 0$.





\end{document}